\RequirePackage{fix-cm}
\documentclass[smallextended]{svjour3}       
\smartqed  
\usepackage{graphicx}
\usepackage{natbib} 
\usepackage{multirow}
\usepackage{booktabs}
\usepackage{array}
\begin{document}

\title{A New Paradigm of Software Service Engineering in the Era of Big Data and Big Service}

\titlerunning{Software Service Engineering in the Era of Big Data and Big Service}        

\author{Xiaofei Xu  \and
	Gianmario Motta \and
	Xianzhi Wang \and
	Zhiying Tu \and
	Hanchuan Xu
}


\institute{Xiaofei Xu \at
	Harbin Institute of Technology, China \at
	\email{xiaofei@hit.edu.cn}           
	\and
	Gianmario Motta \at
	University of Pavia, Italy \at
	\email{gianmario.motta@unipv.it}  
	\and
	Xianzhi Wang \at
	Harbin Institute of Technology, China \at
	\email{xianzhi.wang@hit.edu.cn}
	\and
	Zhiying Tu \at
	Harbin Institute of Technology, China \at
	\email{tzy\_hit@hit.edu.cn}
	\and
	Hanchuan Xu \at
	Harbin Institute of Technology, China \at
	\email{xhc@hit.edu.cn}
}

\date{Received: date / Accepted: date}

\maketitle

\begin{abstract}
Servitization is one of the most significant trends that reshapes the information world and society in recent years.
The requirement of collecting, storing, processing, and sharing of the Big Data has led to massive software resources being developed and made accessible as web-based services to facilitate such process.
These services that handle the Big Data come from various domains and heterogeneous networks, and converge into a huge complicated service network (or ecosystem), called the Big Service.
The key issue facing the big data and big service ecosystem is how to optimally configure and operate the related service resources to serve the specific requirements of possible applications, i.e., how to reuse the existing service resources effectively and efficiently to develop the new applications or software services, to meet the massive individualized requirements of end-users.
Based on analyzing the big service ecosystem, we present in this paper a new paradigm for software service engineering, RE2SEP (Requirement-Engineering Two-Phase of Service Engineering Paradigm), which includes three components: service-oriented requirement engineering, domain-oriented service engineering, and software service development approach.
RE2SEP enables the rapid design and implementation of service solutions to match the requirement propositions of massive individualized customers in the Big Service ecosystem.
A case study on people's mobility service in a smart city environment is given to demonstrate the application of RE2SEP.
RE2SEP can potentially revolutionize the traditional life-cycle oriented software engineering, leading to a new approach to software service engineering.

\end{abstract}

\section{Introduction}
\label{sec:intro}
The past recent years have witnessed drastic shift in information technology. New technologies, such as Web 2.0, cloud computing, Internet of Things (IoT), Internet of Services (IoS), and Big Data, have emerged and influenced human society profoundly.
In particular, Web 2.0 has transformed the World Wide Web from a pool of mere web documents for sharing information into a collaborative network which everyone can contribute to and benefit from.
Recent advancements in Cloud Computing and IoT technologies further bring information technology towards a Big Data era~\cite{Dobre2014}.
Cloud Computing represents an elastic and economical approach for delivering software through the Internet~\cite{Armbrust2010}. With the Everything as a Service (XaaS) architecture, typically including Infrastructure as a Service (IaaS), Platform as a Service (PaaS), and Software as a Service (SaaS) layers, cloud computing provides the networked infrastructure with huge service capacity and enables the virtualization and online consumption of massive services and resources in a cyber-physical world to serve massive customers.
The explosion of social, transactional, and sensor data has strongly influenced the rapid development of Internet of Services (IoS). The services and resources involved in processing the Big Data have dramatically increased in both number and complexity. These services typically come from multi-domains and heterogeneous networks. They are convergent, interrelated, and inter-operated as a huge complicated service network and finally form a service ecosystem, which we call the \textit{Big Service}. 

Formally, we define \textit{Big Service} as the massive and complicated series of services that deal with the Big Data.
Big Service can be considered as a complicated networked business that correlate and exist across the cyber and physical worlds~\cite{xu2015}.
We identify several critical new features of Big Service, namely \textit{Massiveness, Heterogeneity, Complexity, Convergence, Customer focus, Credibility and Value}, which correspond to and enrich the 5Vs features (\textit{Volume, Variety, Velocity, Veracity, Value}) of big data.
A big service ecosystem generally hypothesizes \textit{massive} software services on the Internet, which is also known as IoS. These services show the \textit{complex} forms, come from various sources, and are networked and interoperated across multiple worlds, domains. and \textit{heterogeneous} networks.
The services are gathered, clustered, and finally \textit{converge} into composite services to meet the \textit{customer} requirements.
The basic quality metrics of all services, resources, as well as their delivery approaches, are represented as \textit{credit}, which is used as the basic precondition and foundation for constructing and operating the Big Service. The ultimate goal of Big Service is to create \textit{value}.

With over-abundant data, services, and resources reusable on the Internet, the ecosystem of Big Data and Big Service is changing not only the roles of software and services applications, but also the approaches to their development.
Nowadays, software services have gone far beyond the traditional ``from scratch" development environment and approach.
Modern engineering methods for software service engineering have entered the age of maximally reusing the Internet-accessible open services and sources, such as web services, semantic web, virtualized services, and cloud services, to minimize the development efforts and timespan.
Researchers on software engineering and service engineering are exploring the new paradigms and more efficient approaches for building the software services on-demand of massive individualized customers.
Under this condition, the Big Service perspective provides agility in satisfying massive individualized customer requirements through on-demand application optimization and implementation.

By analyzing the features and phenomena of Big Service, we discover that customer requirements and the development of open, reusable service resources are the keys to developing a successful software service in the Big Data and Big Service ecosystem.
For this reason, this paper focuses on a new paradigm of software service engineering for rapid development of service solutions in the Big Service ecosystem.
This paradigm is called RE2SEP (Requirement-Engineering Two-Phase of Service Engineering Paradigm), involving the new approaches for service requirement engineering and for domain-oriented software service engineering in Big Service ecosystem, respectively.
In particular, requirement engineering presents the requirements and proposition of customers exactly, while service-oriented software engineering focuses on constructing the service solutions based on the reusable service sources to meet those requirements.
Specially, service patterns and features can be identified through the long-term application experiences of services and business in a certain domain. These patterns can then be used for constructing and further optimizing domain service solutions.

This paper is organized as follows: section 1 introduces the background; section 2 discusses the related work; section 3 presents a reference architecture of Big Service and analyzes the Big Service ecosystem; section 4 presents the basic framework of the RE2SEP paradigm; section 5 presents service engineering and approaches in the RE2SEP paradigm, including service-oriented requirement engineering, domain-oriented service engineering, as well as the development approach of software services; section 6 shows a case study of RE2SEP applications; and section 7 gives the conclusion remarks.

\section{Related Work}
\label{sec:related work}
Traditional software engineering processes tpyically follow a recursive top-down approach, where each recursion covers parts of software's entire lifecycle.
For example, the classical \textit{waterfall model} \cite{royce1970} goes from acqisition of high-level user requirements, to software design, and to software implementation and testing to reach workable applications.
Most other process models for software engineering, as summarized by \cite{sommerville2004}, derive recursive or evolutionary procedures from the basic software lifecycle to improve the waterfall model.
For example, the \textit{spiral model} \cite{boehm1988} continuously repeats the waterfall process until a satisfactory result is achieved.

The idea of defining reusable component as a starting point of engineering is first poposed by \textit{Component-Based Software Engineering (CBSE)} and rapidly became prominent in 1968 \cite{mcilroy1968}.
CBSE is a software reuse-based approach that uses software package to encapsulate a relatively independent set of functions or/and related data.
By formally defining and selectively reusing the previously developed components in the same domains, the efficiency and cost of developing different modules of new applications are expected to be notably reduced.
The key distinction of such reuse-based approach is that the components are designed by defining standard interfaces to maximize their reuse and regarded as significant assets to be matched and managed in a dedicated component library.
As a special case of CBSE, the components in \textit{Object Oriented Programming (OOP)} \cite{rumbaugh1991} are specifically objects defined in accordance with people's natural conceptualization of real world entities and interactions.

Another important case of CBSE, \textit{Model Driven Architecture (MDA)} \cite{OMG2011}, was launched by the Object Management Group (OMG) in 2001, with the aim of providing systematic guidelines for structuring specifications as layered models to enable model-driven engineering of software systems.
MDA is a type of forward engineering that produces code from human understandable specifications through model transformation.
The basic layers for transformation involve Computation Independent Model (CIM), Platform Independent Model (PIM), and Platform Specific Model (PSM) from top to the bottom.
CIM focuses on specifying system requirements at the conceptual level, PIM on the details of a system at the system level, and PSM on the details that are specific to the specific development environment and execution platform of the targeted systems.
These models are transformed layer by layer from to the bottom, making the proper model representation and transformation become the key issues of software development in MDA. 

It is worth noting that most of existing engineering methodologies and methods fall into the research field of \textit{Domain Engineering} \cite{harsu2002}, which aims at forming a common infrastructure for reusing the concepts, design, or even implementations of previous software systems within the same application domains.
Domain engineering represents the systematic principles and processes for domain software reuse.
Given that similar systems are repetitively built within a given domain and these systems have many common characteristics despite the differences,
these common characteristics are often regarded as domain knowledge and domain engineering is generally built upon such domain knowledge and would therefore require analyzing and modeling the target domain in advance.
In this sense, domain engineering focuses on a family of systems rather than a single system. 

As the first initiative to the servitization revolution in software engineering via enforcing the textit{Service Oriented Architecture(SOA)}, IBM published the publicly announced the  methodology for implementation SOA in 2004 named \textit{Service-Oriented Modeling and Architecture (SOMA)} \cite{bieberstein2008}.
By extending OOP and CBSE, SOMA covers a broader scope and implements a service-oriented analysis and design (SOAD) process through the identification, specification, and realization of service life cycle.
Components for realizing those services (a.k.a. ``service components") and flows are used to compose services of large granularities.
Web services in SOA are also a type of components, yet have some unique characteristics of services beyond those of ordinary components.
The \textit{Service-Oriented Modeling Framework (SOMF)} \cite{bell2008} is another methodology for software service development.
It is a holistic and anthropomorphic modeling language that employs disciplines and a universal language to provide tactical and strategic solutions to enterprise problems. 
More recently, \textit{Service Model Driven Architecture (SMDA)} \cite{xu2007} apply the MDA approach for developing service systems.
It extends the Unified Modeling language (UML) to define a Unified Service Modeling Language (USML) for describing service models, and focuses on specifying and transforming the multi-dimensional service models and reusing existing services components in developing new service systems.

Based on the combination of the advantages of the previous approaches, this paper presents a new paradigm for Big Service called RE2SEP. In RE2SEP, both the proposition of service requirements and service engineering processes that leverages the domain features are paid special attention to.

\section{Development Environment of Software Services in the Big Service Ecosystem}
\subsection{A Reference Architecture of Big Service}

\begin{figure}[htb]
	\centering
	\includegraphics[scale=0.5]{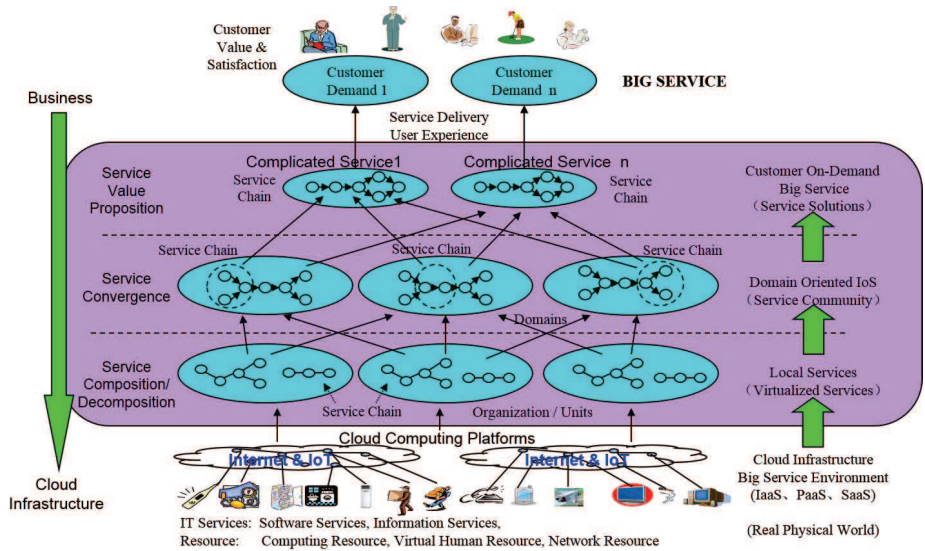}
	\caption{The reference architecture of Big Service}
	\label{fig:The Reference Architecture of Big Service}       
\end{figure}

Big Service is typically constructed through the convergence of services from multi-domains and heterogeneous networks.
A reference architecture of Big Service is shown in Fig.\ref{fig:The Reference Architecture of Big Service} \cite{xu2015}, which consists of three core layers, namely the \textit{local service layer}, \textit{domain-oriented service layer} and \textit{demand-oriented service solution layer}, and two supportive layers, namely \textit{the IT infrastructure layer} at the bottom and\textit{ the client layer} at the top.

In the Big Service ecosystem, the \textit{local service layer} encapsulates the IT infrastructure, including both physical and digital resources, into services through IoT and virtualization.
It further organizes them together with the local services provided by individuals or organizations as the fundamental services for Big Service.
The local services are usually in huge number and often open sources on the Internet. They include either atomic services which are related directly to service providers and real sources, or complex services that are composited by atomic ones.
The services can be further connected or integrated into service chains.
The \textit{domain-oriented service layer} aggregates and composites the local services according to their related business domain, demands and relationships, forming a domain-oriented services communities or IoS.
Services of different granularities are further linked through service chains or service hyper-chains across organizations, domains, and networks to form a complex service ecosystem.
This layer contains a huge number of partial service solutions or service patterns in the service domains and serves as the backbone of the Big Service ecosystem.
The \textit{demand-oriented service solution layer} constructs customized service solutions through the convergence of domain-oriented services to meet the massive individualized customer requirements and to create value.
Finally, the bottom layer, \textit{IT infrastructure layer}, provides the basic conditions and supports for big services, and the topmost layer, \textit{client layer}, delivers services to the customers based on customers' requirements and value propositions.

\subsection{Development Environment of Software in Big Service Ecosystem}

Compared with traditional isolated and self-supported software development environments, software service development environment in the Big Service ecosystem is apparently more efficient, open, flexible, and better organized.

The software services development environment in the Big Service ecosystem is typically based on the massive open software services on the Internet or the cloud.
With the development trends of IT servitization, the software design and development have been also transforming towards a service-orientation.
Under the virtualization philosophy of cloud computing and IoS, all computing infrastructure, platform, software, as well as applications are being developed and published as services.
Moreover, there is an another trend of open APIs from various sources in the industry.
This tread is led by the growing availability of proprietary APIs (such as Sina Weibo API, Tencent API, Amazon Web Services API) and the popularized open APIs for IoT, where various data collected by all types of sensors can be accessed through the Internet.
With these open APIs, the service resources can be invoked and reused by other software services on the Internet.
Nowadays, the available services on the Internet have grown into a huge repository with various sources or service communities.

While most of the services in the Big Service ecosystem are developed or provided by certain organizations, communities, or persons, these services are domain/community-oriented and categorized with some certain communities, e.g., the ProgrammingWeb.com or ApiHub.com.
For both within a single organization and public marketplace on the Internet, the open services forms domain-oriented service communities where services that have similar business purposes, functionalities, or interrelations are classified or even hierarchically structured with business relevance and prior collaboration experience.
Moreover, some common standards for domain APIs are starting to emerge, such as the OpenID standard for cross-border authentication, which further accelerate the above process.
These partially composited domain-oriented services are accumulated into the abundant service sources on the Internet, and become a base to construct more complicated service solutions in the Big Service ecosystem.

From the SOA perspective, the advance of web services technologies as the de facto standards has propelled SOA towards enabling the orchestration of multiple individual services into composite ones.
Service orchestration and choreography represent the two types of service composition technologies that change the software engineering.
Such technologies allow for the maximal reuse of existing software assets in the development of new applications, as well as providing agility to achieve cross-organization interoperability and software refactoring.
The massive diversified services and the composite ones together constitute a huge service pool, where the services can be empirically linked, combined, and clustered according to prior business rules or other concerns (such as functional similarity, business correlation, or data relevance), forming complicated service network or IoS.
Instead of using these services that are previously viewed as separated and distinct in both commercial and technological senses, service convergence happens when software engineering approaches are able to support delivering differentiated or customized applications.
Service convergence opens that possibilities of delivering services across network, platforms, and communities for companies, and of accessing new kinds of comprehensive or value-added services for end-users.

As the service experience accumulate, customers are enabled to present more and more individualized requirements for service systems.
How to construct the adaptive service solutions efficiently and to response the massive individualized customer requirements rapidly become a big challenge.
Also, How to meet the complicated customer requirements with adaptive service solutions becomes a critical factor of software service development and delivery.
All the above challenges call for a new approach of software service development, where people pay more attention to the proposition of customer requirements, the construction of adaptive service solutions by means of reusing open service sources, and the matching of requirements and service solutions.
This new approach is defined as a new paradigm of software service engineering, i.e., RE2SEP, which is described in the following sections.

\section{RE2SEP: A Framework of a New Paradigm of Software Service Engineering}
\subsection{The Framework of RE2SEP}

Software service engineering is derived directly from the concept of software engineering.
It differs from traditional software engineering in emphasizing on the service characteristics and regards service value as the ultimate goal of software systems.
Generally, most of the processes and methodologies in software engineering can be directly used for software service engineering by following the service lifecycle-oriented engineering approach.
For example, the most famous waterfall model specifies guideline for building software from scratch, leading engineers to go sequentially from requirement all the way to software codes by phases.
According to this model, the engineering is consistently working towards service/software implementations from the requirements.
Similar features are possessed by other models such as the spiral model.
Although engineers can iterate the engineering process and repetitive update the engineering results, the basic procedures are always unidirectional and lifecycle oriented.
When compared with traditional software engineering approaches, the MDA focuses on modeling and model transformation from upper-layer (e.g. CIM, PIM) to lower-layer (e.g. PSM), and until the software components, so as to build new software applications rapidly by means of reusing existing software components.
Although the MDA approach has been a dominant paradigm for component-based software development, it is still sometimes low efficient.

From the service-oriented software engineering's perspective, the main task of software service development is to construct and deliver software services that are adaptive to massive individualized customer requirements.
Given the abundant resources of software services and the development trends of software service engineering in the Big Service ecosystem, the software service engineering can be performed efficiently in a bidirectional approach, which includes two phases, namely service-oriented requirement engineering and domain-oriented service engineering.
In the build-time phase of domain-oriented service engineering, the software services are produced by organizations, individuals, and communities, and are accumulated into reusable service resources in a certain domain.
While in the run-time phase of service-oriented requirement engineering, the customer requirement propositions can be analyzed and proposed and then a comprehensive service solution can be constructed by means of convergence and composition of the reusable services from service resources of one or multiple domains.

We propose a new paradigm, namely ``Requirement-Engineering Two-Phase of Service Engineering Paradigm (RE2SEP)", to describe the above approach of software service engineering in the Big Service ecosystem.
Unlike the life-cycle oriented software engineering approaches, RE2SEP is a bidirectional approach with a two-phase process (service oriented requirement engineering and software engineering).
RE2SEP not only reuses the open resources from multi-domains but also matches between service requirements and service solutions using service context as a mediating facility. 

\begin{figure}[htb]
	\centering
	\includegraphics[scale=0.45]{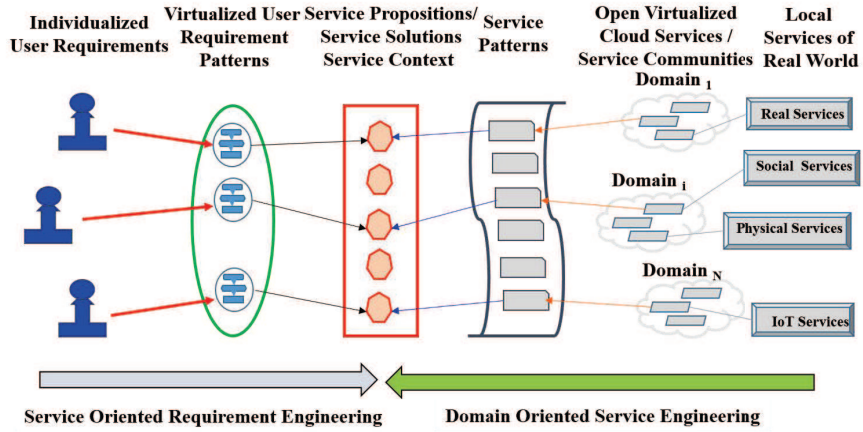}
	\caption{The framework of the RE2SEP Paradigm}
	\label{fig:The framework of the RE2SEP paradigm}     
\end{figure}

The RE2SEP framework is shown in Fig.\ref{fig:The framework of the RE2SEP paradigm}.
It consists of two engineering processes with opposite directions.
The Domain-Oriented Service Engineering (DOSE) process starts from available physical and virtualized software services in a certain domain and domain knowledge, and focuses on how to provide and leverage these services more sufficiently and wisely.
Through virtualization technology, the real physical services (e.g. human services, social services, IoT services, production services,and business services), are virtualized into cloud services and organized as the service resources in the domain-related service community.
The service community contains not only virtualized physical services, but also the networked software services.
Some services can be even gathered from other domains or service communities.
Here, the domain-oriented services normally have certain domain features with domain experience and serve as the carrier of domain knowledge.
In the DOSE process, the services are organized in terms of service patterns to represent typical skeletons of service solutions or sub-solutions.
Further, the service solutions can be constructed to meet the customer requirements, through service convergence from multi-domain service resources as well as service patterns.
The context of service usage is often considered by service providers to deliver the adaptive service solutions to match the individualized customer requirements.

The Service-Oriented Requirement Engineering (SORE) process begins with individualized customer requirements and virtualized requirements description.
It focuses on deriving application-specific service requirement propositions to derive and evaluate the final service solutions.
By analyzing massive individualized customer requirements, SORE is able to extract and define typical requirement patterns that occur frequently in customer requirements.
These requirement patterns are the generic and frequently occurring representation of fragments or pieces of user requirements in a certain application domain.
They can be generated based on the experienced applications or business processes, and can be easily reused to form the customer requirement propositions. %
For every individual customer, the user requirement proposition would be related to a certain context of service usage which contains requirement patterns.
Therefore, the link of the user requirement propositions and the service solutions becomes the key interaction point of the two phases of RE2SEP.
To obtain the best matching of user requirement and service solutions, the matching mechanism between requirement patterns/requirements and service patterns/services can be facilitated by means of service contexts.
If requirement patterns and service patterns can both be matched to some service contexts, then a mapping can be set between those patterns.
The mapping of requirement patterns and service patterns can be used for optimizing service solutions.
Further, we can select and compose the service patterns to form the final service solutions, which map the customer requirement with more corresponding requirement patterns.

To exemplify, the philosophy of the RE2SEP can be regarded as analogical to the medical treatment process, where SORE and DOSE are analogical to the diagnosis process in the hospital and the medicine production process by medicine companies, respectively.
In the medical treatment process, a patient (analogical to a user/customer) with personal physical conditions (analogical to individualized user requirements) can be diagnosed to have some diseases (analogical to requirement patterns).
For each disease, there are different options of usage solutions of medicines (service patterns).
Each solution may contain multiple types of drugs or medicines (available services), which are produced from drug materials (atomic services) by different medicine companies (service providers).
The ultimate goal of the medical treatment process is to provide an adaptive prescription (service solution) to shoot the target spots for healing the diseases of the patient, considering comprehensive the multiple conditions (service context) of the patient, such as the severity of disease of the patient, the prices of the medicines, and the conflicts among the medicines.
There are two phases in the above medical treatment process---the hospital diagnosis process and the medicine company production process.
The doctors in hospitals pay little attention to the details of medicine production processes, while the medicine companies pay more attention to the healed diseases. Similar features can be found in the RE2SEP paradigm. 

\subsection{Key Concepts in the RE2SEP Paradigm}

In order to explain and understand the principle of the RE2EP paradigm, we define several key concepts as follows.

(1) \textbf{\textit{Service requirement proposition}} presents customer's constraints and expectations for the final service solutions regarding the business functionality, service performance, and value.
The description should have sufficient expression ability to describe various requirements of different customers. There are many requirement models and methods which can be used to define service requirement proposition.

(2) \textbf{\textit{Service requirement pattern}} is a modularized piece of description of customers' service requirements, which represents a common routine or relatively stable service process or sub-solution in domain business services.
The service requirement patterns can be presented using requirement models, and are expected to be reused in satisfying different customer requirements in certain domains. The usage experience in an application domain is often used to define service requirement patterns.

(3) \textbf{\textit{Service context}} refers to the information about the condition under which a service is applicable to satisfying a specific service requirement pattern.
It can be used for mapping service requirements and service solutions in a fine granularity.
Service context is often related to application scenarios where environmental information or customer's status may affect the matching.

(4) \textbf{\textit{Service solutions}} are the deliverable integrated service sets for satisfying specific customer requirements. Service solutions are targeted directly at the customer's service requirements and the ultimate output of service engineering.
A service solution can be constructed and delivered through a convergence of multiple services and service patterns from one or multiple domains. There are many service models and methods which can be used to define service solutions.

(5) \textbf{\textit{Service chains/hyper-chains/sub-chains}} refer to the linkages and correlations among multiple services or service patterns used to form more complex services, so as to support the construction of service solutions. Generally speaking, the linkages between basic services or within a single service pattern are called service chains. A piece of partial service chain is called service sub-chain. The linkages among multiple service patterns or composite services across multi-domains are called service hyper-chains.

(6) \textbf{\textit{Service pattern}} is the relatively independent business unit describing service processes and related services and resources. Service pattern refers to a typical complete or partial service solution in a certain domain. Service patterns are often defined based on the application experience of business services in a domain, which are the carriers of prior knowledge in domain-oriented service engineering experiences. Service patterns are analogical to phrases in sentences and can be instantiated for execution. 

(7) \textbf{\textit{Open virtualized services}} refer to the services that are published and provided by different service providers from different organizations, communities, domains, networks, individuals and physical locations, and maintained in the cloud as open resources. They can be clustered, selected and combined for both management and usage purposes in the RE2SEP.

(8) \textbf{\textit{Domain services}} refer to the services that are developed and published by different vendors, organizations, and clustered into a certain business service community/sector in a certain domain. These services are the fundamental building blocks of Big Service. 

\section{Approaches of Software Service Engineering in the RE2SEP Paradigm}
\subsection{Service Oriented Requirement Engineering in the RE2SEP Paradigm}

\begin{figure}[htb]
	\centering
	\includegraphics[scale=0.45]{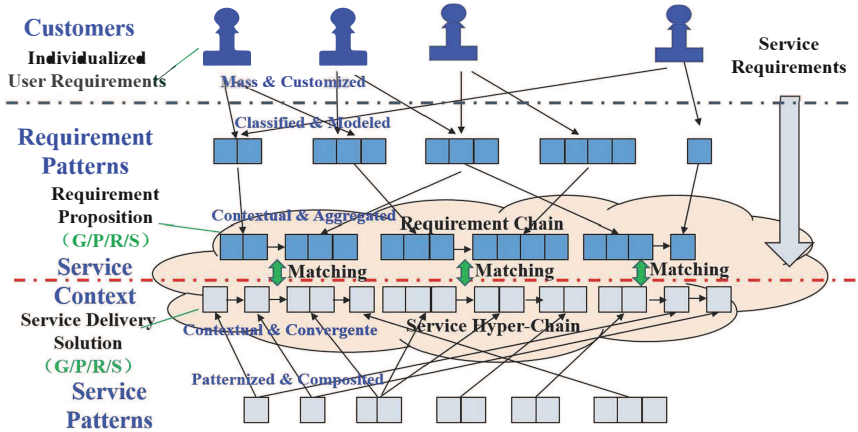}
	\caption{Architecture of Service Oriented Requirement Engineering in the RE2SEP Paradigm}
	\label{fig:Architecture of Service Oriented Requirement Engineering in the RE2SEP Paradigm}     
\end{figure}

The architecture of Service Oriented Requirement Engineering (SORE) in the RE2SEP Paradigm is shown in Fig.\ref{fig:Architecture of Service Oriented Requirement Engineering in the RE2SEP Paradigm}. The main objective of SORE is to acquire customers' massive and individualized requirements and to describe the service requirement propositions. SORE facilitates the representation and processing of those requirements to enable effective service solutions recommendation.

SORE is a top-down approach where user requirements are initially presented by massive and individualized customers. These customer requirements are then represented and modeled using service requirement models.
By analyzing various customer requirements, the requirements can be classified and clustered according to the related business and the similarity in their specific specifications.
Typical non-functional specification of service requirements includes those related to Quality of Service(QoS), constraints, customer-defined rules, customers' preferences and value proposition regarding the expected service solutions.
Customers' value proposition specifies how different aspects affect the final customer satisfaction.
A service requirement proposition can be represented as RGPS meta-models \cite{hu2010}, i.e., the aspects of the goal, processes, roles, services of the service requirement and the expected service solution.

In order to improve the matching service solutions to service requirements, it is expected that the service requirement proposition should be precise and measurable.
Since service requirements in the same domain often share similar contents, the typical fragments in the modularized requirement model can be extracted and well-organized as prior knowledge.
Therefore, service requirement patterns are used to represent the modularized pieces of the description of customer requirements including the usage experience of domain application, according to the business usage context.
Service requirement patterns often share a similar or identical mapping in a pool of services patterns in the service engineering process.
While a new requirement can be represented as several relevant service requirement patterns, the massive individualized requirements of customers can be aggregated into limited requirement patterns.
Through selecting and aggregating multiple service requirement patterns, a specific requirement proposition can be presented precisely based on the service context of related application domains.
Service context is be often used on this basis to match service solutions to service requirements.

\subsection{Domain Oriented Service Engineering in the RE2SEP Paradigm}
\begin{figure}[htb]
    \centering
    \includegraphics[scale=0.45]{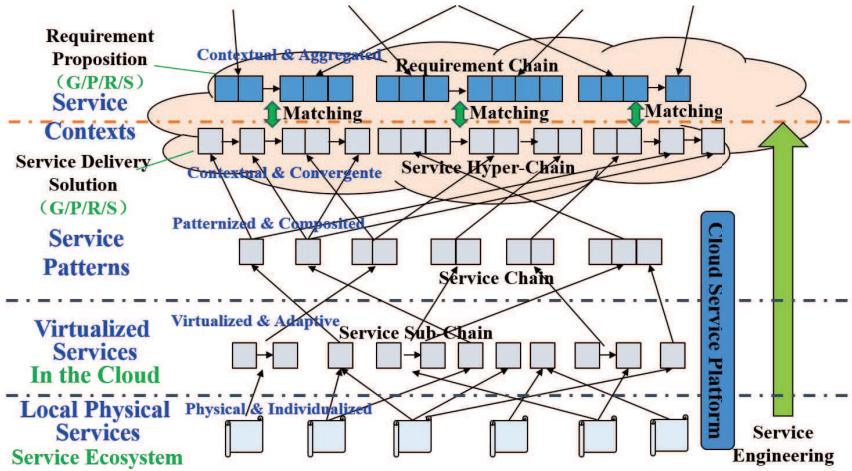}
    \caption{Architecture of Domain Oriented Service Engineering in the RE2SEP Paradigm}
    \label{fig:Architecture of Domain Oriented Service Engineering in the RE2SEP Paradigm}     
\end{figure}

The Domain Oriented Service Engineering (DOSE) is the important part of the RE2SEP Paradigm. Similar to the architecture of Big Service, the DOSE architecture consists of four conceptual layers, i.e. physical service layer, virtualized service layer, service pattern layer and service solution layer, as shown in Fig.\ref{fig:Architecture of Domain Oriented Service Engineering in the RE2SEP Paradigm}. The physical services are mostly individualized, which form the fundamental basis of the service ecosystem. They could be virtualized and adapted into the cloud virtualized services. The virtualized services are firstly related to certain organizations or developers, with the domain features such as apriority, correlativity, similarity, and domain knowledge contained in historical service experiences. These virtualized services and the networked services can be used to form the open source on the Internet. As in a middle service entity layer, service pattern represents a form of such formalization results which aggregate and composite different virtualized services to represent typical service sub-solutions or solution fragments. Generally, the service patterns are related to a certain domain. According to the given service usage context, the service patterns are further linked together by means of service hyper-chain to form final service solutions. The DOSE is a bottom-up approach based on service patterns and open service sources and is often facilitated by an integrated cloud service platform. The DOSE procedure deals with the handling of fundamental services, management of virtualized services, construction of service chains and sub-chains, services patterns and service hyper-chains or service community networks, and formation of service solutions.

The basic services of domains are provided by individuals or organizations. These service providers are typically based on different technological platforms and under heterogeneous network conditions. Most of the basic services are initially developed and accessed locally within a single organization's business domain. Many services are connected with physical services and even performed in the real world of the domain. Once these services come into open source or are accessible globally, they become the primary building blocks of the Big Service ecosystem.

In DOSE, the virtualized services are encapsulated, adaptive, classified, interrelated, and prioritized to form well-organized service communities in the cloud.
Virtualization enables the distributed domain services in the real world to be published using standard interfaces, or deployed to some centralized platforms to be queried and invoked directly.
The virtualized services should be classified based on their related domains and business functionalities.
The services with explicit business correlations are linked together.
In order to use the services efficiently, the services are further prioritized and adaptive based on customer's comments and performance in usage history. The services frequently used together in historical experiences are leveraged as prior knowledge as well.

The linked basic services form the fundamental service sub-chains, which focus on the correlations in business and usage history of the services; while service patterns focus more on the reusability of these structures.
Service pattern can be either defined by domain experts or discovered automatically through data mining methods.
Different from generic virtualized services with service sub-chains,service patterns can be either typical independent business processes or a partial service solutions in the domain, and are more solution-oriented and business application-oriented. 
Service patterns can be considered as relatively independent business units that are easier to be reused for constructing service solutions.
The difference between service pattern and basic service can be analogical to the difference between molecule/ molecular group and atom.

Similar to the linkages among services within a single service pattern, service chain is a higher-level concept of service sub-chain.
In satisfying customer's requirements at the application level,
service solutions can be more precisely and efficiently constructed through the composition of multiple service patterns as well as services from multiple domains. 
The service patterns are further converged and linked together by service hyper-chain according to the application service context. 

Service solutions represent the final deliverables to meet the customer requirements and are obtained based on context-aware service composition and the convergence from multi-sourced service resources (e.g., physical services, real world resources, cloud services, IoT services, IoS services, as well as different network protocols).
In the Big Service ecosystem, the services for solutions may come from multiple domains, networks, and worlds.
In order to match the customer requirements precisely with service solutions, the mapping correlations among service patterns, service requirement patterns, requirement proposition, service solution, as well as service context, should be analyzed. 

In the RE2SEP paradigm, various factors, including the open software service resources, domain knowledge, and prior knowledge about services' usage history play very important roles in facilitating the software engineering process.
The abundant reusable services and software resources are preconditions of the RE2SEP paradigm as they form the fundamental basis of the Big Service ecosystem.
Only with sufficient available services, can domain features be revealed in the service engineering process.
The abundant services also make it possible to define and reuse the prior knowledge about a specific domain or type of applications, thus improving the engineering efficiency and the quality of service solutions.
In order to reuse the abundant open service resources in one and multi-domains, the virtualized services should be well structured and integrated into service patterns through service sub-chains, service chains and service hyper-chains in the Big Service ecosystem.

\subsection{The Approach to the Development of Software Services in the RE2SEP Paradigm}

As aforementioned, the RE2SEP paradigm follows a bidirectional two-phase process, i.e. SORE and DOSE.
Correspondingly, the development approach of software services in the RE2SEP paradigm exhibits as a two-phase procedure, i.e., the build-time phase and the run-time phase.
The build-time phase mainly concerns the preparation of development environments and resources and services to facilitate service requirement specification.
The run-time phase deals with dynamical service construction and execution according to the massive individualized customer requirements.
It includes three tasks: customer requirements specification based on requirement patterns, service convergence based on service patterns, and service requirement-solution matching.
The development procedure of software services in the RE2SEP paradigm consists of five steps as follows:

(1)    \textbf{Construction of service development environment in the cloud platforms}

Preparation of service development environment includes the facility for gathering and organizing real world services (such as social services, physical services, and IoT services) from one or multi-domains onto the cloud platform, virtualizing services, defining and structuring service patterns, etc. The physical services will be virtualized onto the cloud platforms, and classified depending on where and how they will be used. The prior knowledge of services' usage in the domains is often used to guide the development of new services or the design of new service patterns. The services are further improved or restricted by business rules and domain constraints. The services, which are configured, deployed, and published in the cloud platforms, are prepared and organized based on the experienced service solutions in the past and used in the future service solutions. The domain related open service resources is the key part of the service development environment. The service development environment should also support defining and publishing new services in case that no existing service can satisfy the requirements.

(2)    \textbf{Presenting service requirements based on service requirement patterns}

The models and structure of service requirements should be presented to describe different individualized customer requirements. The service requirement propositions can be transformed into presentations composed by requirement patterns. It can be found that even various massive customers have relatively limited types of service requirements which are the composition of requirement patterns formed through previous usage experience in the certain domains. The service requirement patterns are designed based on typical business usage manners in related certain domains, which can be easily matched by certain service patterns of the final service solutions. Some rules should be defined on which fragments of customer requirement can be replaced by requirement patterns, and how to deal with overlaps between different service requirement patterns.

(3)    \textbf{Service convergence based on open services resource and service patterns}

Though service patterns are designed for providing partial service solutions, they may not fit service requirements or requirement patterns directly and exactly in real applications. Through service convergence of different services or service patterns, the service solutions can be formed to fulfill the service requirements. Service convergence requires the compatibility among different services and service patterns from different sources, domains, organizations, and networks. Service convergence is targeted straightly at the typical solutions of service applications and is aimed at providing some prepared composite service sets for typical solutions or sub-solutions directly based on the available services and service patterns in the service development environment before a specific actual requirement is known. Service chains and hyper-chains are designed to facilitate these typical solutions or sub-solutions consisting of the converged service sets. It is very often that the service solutions are constructed through the convergence of service patterns and services implemented by open service sources on the Internet or IoS.

(4)    \textbf{Matching service requirements with service solutions}

This step concerns matching between service requirements and service solutions at different granularities and aspects. A service solution could either be identified by selecting/modifying existing prepared services sets, or by combining existing/modified service patterns as newly constructed results. Service requirement patterns are matched with service patterns with service context as mediation. Service context defines service usage information at different levels, such as how different requirement patterns and services patterns are mapped to each other, and how different service requirements can be satisfied by different available prepared service sets. A GPRS (Goal-Process-Role-Service) requirement meta-model \cite{hu2010} can be used for facilitating the matching of service requirements and solutions. The goal, processes, roles, and services are the key factors of both requirements and solutions of services. Through matching these factors, the service solutions with service patterns will get the consistent fulfillment of service requirements with requirement patterns.

For a certain service system, the result of matching can also be evaluated in five aspects, i.e., matching efficiency, matching precision, matching effect, QoS of solutions, and scopes of matching choice. The matching efficiency measures the response time for the producing new matching results. The matching precision indicates the consistency between the outcome and the customer's expectations. The matching effect expresses the benefits and value of service solutions. The QoS of service solution focuses on the service quality factors in the matched service solutions and the satisfaction of customers. The scopes of matching choice stand for the abundance of alternative solutions matching to different service requirements.

(5)    \textbf{Service delivery and execution}

The optimized service solutions would be delivered through the Internet of Services with the support of open service sources. In the Big Service ecosystem, the service delivery requires a cloud-based service running platform, a service center and service library, web portals or mobile user-end for interaction with customers. The service process planning, resource scheduling, and service execution are performed by a service execution engine on the cloud platform. The service center is responsible for the service resource management, service invoking and coordination, service monitoring related to service delivery, and the web user-interfaces. Some services may be also delivered by humanity and physical entities connected with IoT and IoS. The service delivery approaches give the important influences on QoS and satisfaction of customers. The customers can give feedbacks and comments on the delivered service solutions, which would be used as evaluation results from the customer side. The performance and quality of services and service patterns may be monitored and evaluated during or after service execution. The feedback would be useful for improvement of Big Service ecosystem in the future.

\section{IRMA Project: a Case Study of RE2SEP}
In order to understand the principles of the RE2SEP, a case study called IRMA (Integrated Real-time Mobility Assistant), is introduced, whose architecture and development are close to the RE2SEP framework. IRMA, a project on mobility services in a smart city environment, has been developed by the Lab of Service Engineering of University of Pavia, Italy \cite{motta2013}, and tested on the city of Pavia in cooperation of the Municipality of Pavia.

\subsection{Overview and Architecture of IRMA}
\begin{figure}[htb]
    \centering
    \includegraphics{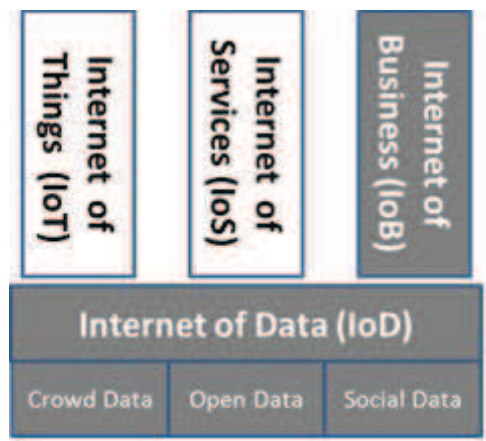}
    \caption{Layers of IRMA}
    \label{fig:Layers of IRMA}     
\end{figure}

IRMA was born with the idea of providing users a tool for managing their mobility on any combination of public / shared transports – bus, underground, train, plane, taxi. This implied not only to put together static information as maps and timetables, but also event information as the real-time position of transports, delays, alerts issued by transport authorities, and, last but not the least, tweets from the social network. This wide information ecosystem includes not only IoT, wherefrom real-time position of transport comes, but also IoS that provides services as maps, timetables, overall information and alike. The ecosystem of services and information had to serve the ecosystem of urban stakeholders. This includes (a) individual users (as citizens and tourist), who can use any available transport system, (b) impeded users, who can use only some transports on some itineraries (as elderly, wheel-chaired, and blind people) (c) transport providers and authorities (as public transport, taxi, shared transport providers, and municipalities) who manage and operate the transports. In short, IRMA, or in another name, Integrated Real-time Mobility Information and Services or IRMIS, is a layered set of services, ranging from IoT to IoS to the complicated services propositions, that, for the sake of uniformity, we call Internet of Business (IoB). In turn, IoD provides access services to Internet data sources (as Fig.\ref{fig:Layers of IRMA} shows). 

\subsection{RE2SEP case description for trip planner of IRMA }

\begin{figure}[htb]
    \centering
    \includegraphics[scale=0.28]{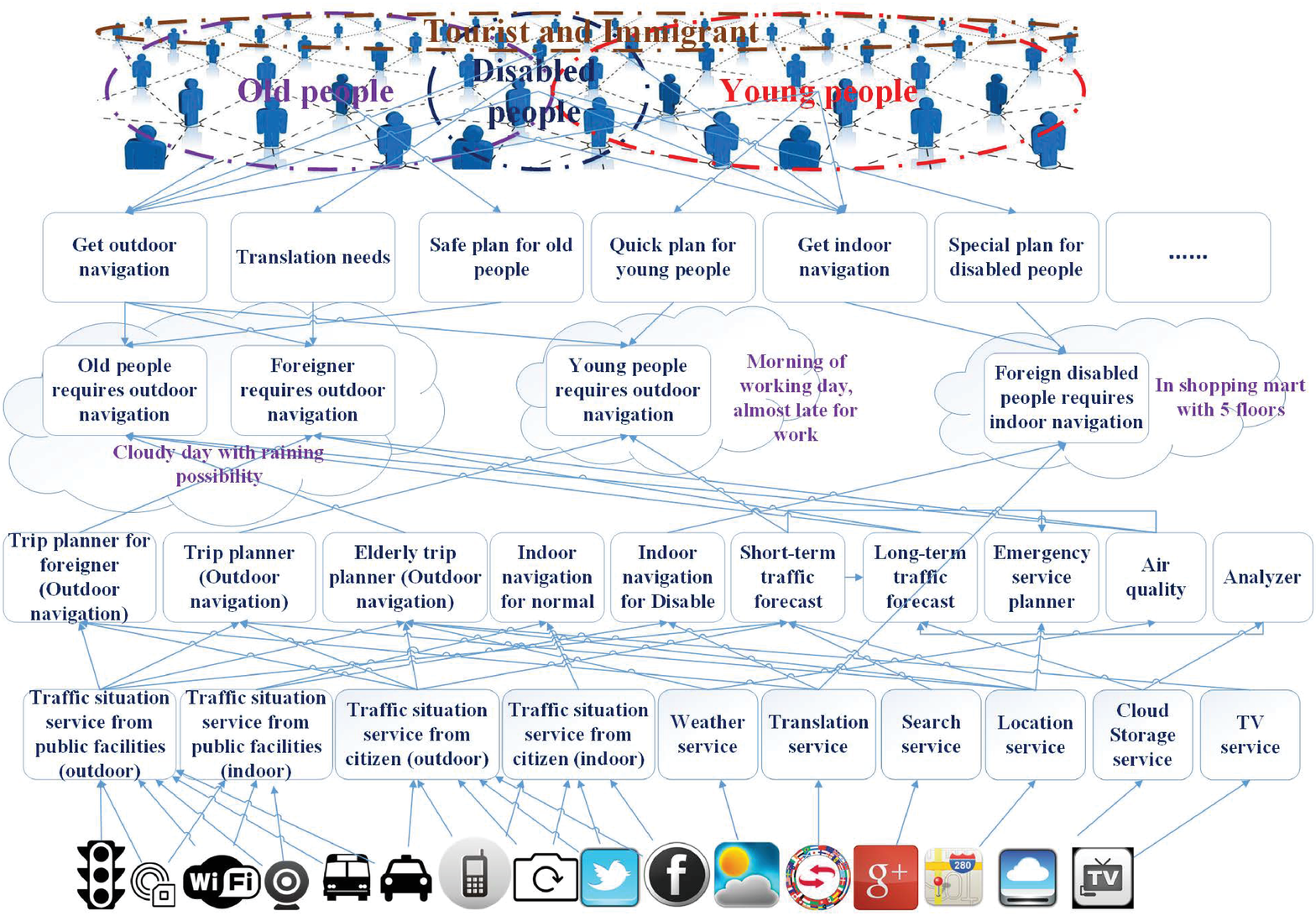}
    \caption{Case Description of RE2SEP in IRMA}
    \label{fig:Case description of RE2SEP in IRMA}     
\end{figure}

One objective of IRMA project is to provide mobility to everyone everywhere, thus including also disabled people, and both outdoor and indoor mobility inside complex buildings. It defines a platform that fits small cities (as Pavia, with 70,000 inhabitants and 2,000 students), midsize cities (500,000 inhabitants) and large metropolitan areas with complex transport system (underground, buses, tramways, taxi, shared vehicles). Fig.\ref{fig:Case description of RE2SEP in IRMA} demonstrates a case description of RE2SEP for trip planners in IRMA.

There are numerous users with various kinds of individual requirements, thus the platform must be able to cover these requirements as many as possible. Because of different natural attributes (gender, age, nation, and etc.) and hobbies or interests, people might have diverse behaviors and needs. By grouping these users and refining these requirements, some common/similar requirement fragments can be extracted. Part of these requirement fragments could be common for all, and part of them could only belong to a specific user group. For example, the requirement fragment ``Get outdoor navigation" in fig. 6 is the public demand of everybody who needs a routing plan; in contrast, ``Translation needs" is special service for foreign visitors who do not understand native language. For the old people, they are not in a rush, so the safer trip plan is better; in contrast, young people always require the short path. To combine the common requirement fragment with the specific one, we can acquire one requirement pattern for one specific user group. For example, the ``Old people requires outdoor navigation", ``Young people requires outdoor navigation", and ``Foreigner requires outdoor navigation".

\begin{table}[htb]
    \centering
    \caption{List of IRMA/IRMIS service patterns}
    \begin{tabular}{p{2cm}p{2.5cm}p{6cm}}
        \toprule
        Category & Service & Service description \\
        \midrule
        \multirow{3}[2]{2cm} & Trip Planner (Outdoor navigation) & It supports end to end trips in urban areas; it covers the whole itinerary cycle on public and shared transport systems; it defines, tracks real-time and stores itineraries; it incorporates POIs (Point Of Interest)  \\
         
        Services to Individuals& Indoor navigation & It targets navigation inside complex buildings, with a logic similar to the Urban Trip Planner. It transforms indoor maps in OSM, uses a specific orientation technology based on Wi-Fi anchors and provides a network of POIs that are visually described. \\
        & Alert  & Alert and real-time mobility rescheduling in front of unexpected events. It links and filters social, web, crowd, collaborative and sensor data.  \\
        \midrule
        \multirow{2}[2]{2cm}& Elderly  & A version of Trip Planner for older people that  can run also on smart TV, with usage of avatar animated menus, specific additional services as map of POIs , easy transports, safe ways, etc. \\
        Social Services& Disabled  & It addresses both blind and wheel chaired people. It guides disabled people on a barrier-less itinerary. It includes indoor mobility, the position of disabled people, healthcare services and transport provider. It implies the development of a smart stick with a precision around 1 cm.   \\
        \midrule
        \multirow{5}[2]{2cm}{} & Analyzer & Analysis of flows of public transport, private vehicles and people (time series and real time); it links multiple data sources: social networks, crowd, collaborative applications, mobile network, transit sensors, noise sensors; it provides input to long and short term planning  \\
        & Short-term traffic forecast  & It predicts the forthcoming traffic patterns (e.g. next half hour) in order to detect possible disruptions of mobility patterns \\
        Services to Municipalities& Long-term traffic forecast & Long term traffic forecast service to predict the impact on the city traffic implied by long-term modifications of the traffic flows in the area and also by planned events (shows, fairs, etc.)  \\
        & Emergency service planner & Planning/real-time rescheduling of mobility in front of unexpected events; uses data from multiple nodes of a transit network for enhancing reactive mobility services.  \\
        & Air quality & It predicts the impact on air quality suffered by the neighbors due to construction activities and changes in traffic flows. \\
        \bottomrule
    \end{tabular}%
    \label{tab:List of IRMA/IRMIS service patterns}%
\end{table}%

Meanwhile, there are abundant physical services that have been prepared to fulfill the massive public requirements. As mentioned, through the IoD layer, the IRMA platform can access the IoT devices, such as vehicles, traffic lights, WiFi sensors, noise sensors that measure traffic and any device that collects data on the mobility of vehicles or people. Also, it can obtain the IoS services, i.e. web and mobile applications that provide information (e.g. maps, translation, and search) and social networks (e.g. Google, Twitter etc.). In addition, mobility information is to be accessed by a variety of devices, that include Android and web platforms, as well as special devices for disabled (smart stick) and elderly people (smart TV), thus enabling an easy and pervasive cross-platform access to all stakeholders and also an easy portability. In short, IoD layer converges all these physical services and IoS services into the cloud and make them become accessible virtualized services. Some of these services have already been mashed as a new service, or can be combined to form a new service sub-chain. For example, ``Traffic situation service from public facilities (outdoor)" mashes several physical services from IoT devices, such as vehicles, traffic lights, WiFi sensors, cameras, and bus. ``Traffic situation service from a citizen (indoor)" mashes several IoS services and physical services, such as information from the social network, personal cameras, and mobile devices. In practice, these combined services are always called/requested together to solve common/similar user requirements/requirement patterns. Thus, they can be cheated as one---service pattern. Table 1 shows a list of IRMA/IRMIS service patterns.

Specific requirement proposition is always proposed in specially appoint context. Normally, this requirement proposition is composed of one or more requirement patterns. For example, requirement proposition -``Old people requires outdoor navigation" consists of ``Get outdoor navigation" pattern and ``Safe plan for old people" pattern. Under the constraints led by this context, the requirement proposition becomes more concrete. For instance, an old man has a travel plan on a cloudy day with raining possibility. Then, ``Weather forecast" requirement would be necessary. With the help of these constraints, these requirements/requirement patterns can be matched with the most suitable service patterns. Such as, service patterns, ``elderly trip planner (Outdoor navigation)", ``long-term traffic forecast", and ``weather service", will be called simultaneously to serve an old man to make a safe travel schedule in the bad weather. In contrast, if a young people is almost late for work in a morning of working day, even though it is a rainy day, he will never care about the weather. He needs the service patterns, ``short-term traffic forecast" and ``real-time outdoor navigation", badly to find the fastest trip plan with less traffic jam so that he can arrive his company in time.

\subsection{Design and Implementation of IRMA}
\begin{figure}[htb]
    \centering
    \includegraphics[scale=0.9]{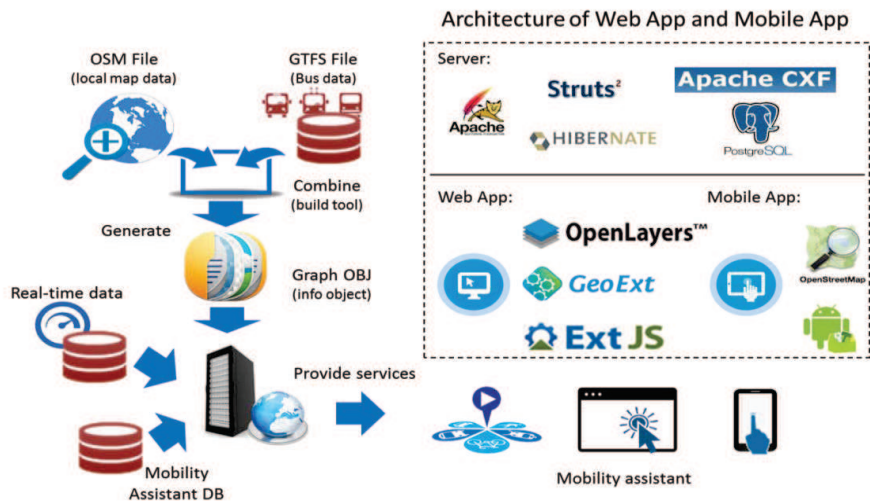}
    \caption{The hierarchy of the Trip Planner service in the IRMA project}
    \label{fig:the hierarchy of the Trip Planner service in the IRMA project}     
\end{figure}

The Fig.\ref{fig:the hierarchy of the Trip Planner service in the IRMA project}  details the layers of services and information in an oversimplified Trip Planner, in turn, the simplest service proposition of IRMA. Trip planner, of course, includes Web and Android applications. The figure shows the stack of Open Source platforms that build the proposition, specifically: 
\begin{itemize}
    \item[-] OSM (Open Street Map) that publishes map data;
\item[-] Timetable, coded in GTFS, a template that describes relevant properties of transports;
\item[-] Real-time data on the position of buses, that are a published by the local Transport Authority, also coded according to a GTFS template for real time;
\item[-] The composited service which combines static and real-time data, that is customer developed; 
\item[-] The value proposition for citizens that runs on web and Android.
\end{itemize}

From Fig.\ref{fig:Case description of RE2SEP in IRMA} and Fig.\ref{fig:the hierarchy of the Trip Planner service in the IRMA project} , one can be easily elicited how the top down - bottom up approach works. At the outset, the designer shall understand the needs of the individual who wants to move around on public transport. First, he shall model the needs and related information requirements – choose the transport, travel, capture alerts etc. – by Use Case diagrams, Goal Oriented diagrams or similar high-level techniques. Then he or she shall define a model that is computer oriented, but still independent from the implementation platform, thus identifying the classes and sequence diagrams or other UML diagrams. At this point, he or she should search what kind of information sources and available services of the lower level he or she can use on the internet, and then define the model that is specific to the platform, that in IRMA are all Opens Sources. Hence the analysis cycle goes through stages that are alike Computation Independent Model (CIM), Platform Independent Model (PIM), and Platform Specific Model (PSM).  

\section{Conclusion}
Servitization is becoming one of the important trends of IT in the big data and Big Service times, where most softwares are developed in the format of services and become open resources on the Internet of Services.
The abundant open reusable service resources from multiple domains and heterogeneous networks provide us a chance of applying these services to develop new applications or software services more rapidly, with the aim of satisfying massive individualized customer requirements.
This paper presents a new paradigm of software service engineering, RE2SEP, which includes service-oriented requirement engineering, domain-oriented service engineering, and development approaches of software services as three phases.
RE2SEP introduces several new concepts, such as requirement patterns and service patterns, for matching customer requirements with adaptive service solutions effectively and efficiently.
RE2SEP builds upon open resources and design and implement adaptive service solutions efficiently to meet the individualized customer requirements in Big Service ecosystem.
IRMA, a case study of RE2SEP application, is given to demonstrate the advantages of this approach.
With the rapid development and popularization of the Big Service ecosystem, the RE2SEP paradigm is expected to lead a new approach to software service engineering.

Our further work involves working on the detailed service engineering and methodology, including normalization of RE2SEP paradigm, service modelling, service pattern description, RE2SEP service solution engineering, service requirement analysis and modelling, service requirement pattern definition, service solution/service requirement matching, RE2SEP methodology and guideline, RE2SEP support tools and so on.
More application case studies will also be conduced.


\bibliographystyle{plain}
\bibliography{hanchuan}   

%
%

\end{document}